\pdfoutput=1 %%%for arxiv to accept pdflatex. 好像不加也能自动识别。本地编译后，整个目录压缩成zip，然后上传后会自动解压缩，一次搞定！

\documentclass[conference]{IEEEtran}
\IEEEoverridecommandlockouts
\usepackage{amsmath}

\usepackage{amsopn}
\DeclareMathOperator{\diag}{diag}

\usepackage{float}

\usepackage{amssymb}

\usepackage{amsfonts}

\usepackage{amsthm}

\usepackage{graphicx}
\usepackage{booktabs}

\usepackage{caption3} % load caption package kernel first
\DeclareCaptionOption{parskip}[]{} % disable "parskip" caption option
\usepackage[small]{caption}

\usepackage{setspace,graphicx,multirow}%,subfigure
\usepackage{subcaption}

\usepackage{algorithm}
\usepackage[]{algpseudocode}
\makeatletter
\def\BState{\State\hskip-\ALG@thistlm}
\makeatother
\usepackage{algcompatible}

\newcommand*\diff{\mathop{}\!\mathrm{d}}

\usepackage{float}

\usepackage{cite}

\usepackage{fancyhdr} % copyright

\usepackage{color}
\ifodd 1
\newcommand{\com}[1]{\textbf{\color{blue} (COMMENT: #1)}} %comment of the text
\else

\newcommand{\com}[1]{}
\fi

\usepackage{enumitem}
\setenumerate[1]{itemsep=0pt,partopsep=0pt,parsep=\parskip,topsep=0pt}
\setitemize[1]{itemsep=0pt,partopsep=0pt,parsep=\parskip,topsep=0pt}
\setdescription{itemsep=0pt,partopsep=0pt,parsep=\parskip,topsep=0pt}

\usepackage{verbatim}
\begin{document}
\bibliographystyle{IEEEtran}
\bstctlcite{IEEEexample:BSTcontrol}

\title{Site-Specific Deployment Optimization of Intelligent Reflecting Surface for Coverage Enhancement}

\author{Dongsheng~Fu, Xintong~Chen, Jiangbin~Lyu,~\IEEEmembership{Member,~IEEE}, and~Liqun~Fu,~\IEEEmembership{Senior Member,~IEEE}%

\thanks{This work was supported in part by the Natural Science Foundation of Fujian Province (No. 2023J01002), the Natural Science Fundation of Xiamen (No. 3502Z202372002), the Guangdong Basic and Applied Basic Research Foundation (No. 2023A1515030216), the National Natural Science Foundation of China (No. U23A20281, No. 61801408), and the Fundamental Research Funds for the Central Universities (No. 20720220078). }%
\thanks{The authors are with the School of Informatics, Xiamen University, China, and also with the Shenzhen Research Institute of Xiamen University, China (email: \{dongshengfu, xintongchen\}@stu.xmu.edu.cn; \{ljb, liqun\}@xmu.edu.cn). \textit{Corresponding author: Jiangbin Lyu}.}}
%and Key Laboratory of Underwater Acoustic Communication and Marine Information Technology (Ministry of Education), 
%\date{}

\maketitle
\thispagestyle{fancy}
% \lfoot{979-8-3503-9973-8/23/\$31.00~\copyright~2023 IEEE}
\cfoot{}
\renewcommand{\headrulewidth}{0mm}
\begin{abstract}
    Intelligent Reflecting Surface (IRS) is a promising technology for next generation wireless networks. Despite substantial research in IRS-aided communications, the assumed antenna and channel models are typically simplified without considering site-specific characteristics, which in turn critically affect the IRS deployment and performance in a given environment.
    In this paper, we first investigate the link-level performance of active or passive IRS taking into account the IRS element radiation pattern (ERP) as well as the antenna radiation pattern of the access point (AP). Then the network-level coverage performance is evaluated/optimized in site-specific multi-building scenarios, by properly deploying multiple IRSs on candidate building facets to serve a given set of users or Points of Interests (PoIs). The problem is reduced to an integer linear programming (ILP) based on given link-level metrics, which is then solved efficiently under moderate network sizes. 
    Numerical results confirm the impact of AP antenna/IRS element pattern on the link-level performance. In addition, it is found that active IRSs, though associated with higher hardware complexity and cost, significantly improve the site-specific network coverage performance in terms of average ergodic rate and fairness among the PoIs as well as the range of serving area, compared with passive IRSs that have a much larger number of elements.
\end{abstract}

% \keywords{Deployment, Active IRS, Lagrangian Relaxation, Heuristic, Throughput, Radiomap}
% \keywords{Site-specific Deployment, Active IRS, Passive IRS, Ergodic throughput, Coverage Enhancement}

\begin{IEEEkeywords}
Site-specific Deployment, Active IRS, Passive IRS, Ergodic Throughput, Coverage Enhancement
\end{IEEEkeywords}

\section{Introduction}

Intelligent Reflecting Surface (IRS) is a promising new paradigm capable of reconfiguring the radio propagation environment by controlling a large number of low-cost reflective elements \cite{wu2019towards}. 
%One of the key application of IRS technology is the improvement of signal channels, the expansion of network coverage, the reduction of blind spots, and the enhancement of communication security\cite{wu2021intelligent}.
%This transformative technology enriches the connection between transmitter (TX) and receiver (RX), opening up a world of possibilities in wireless communication networks.
The IRS-aided communication paradigm has attracted intensive research studies on its hardware architecture\cite{dai2020reconfigurable}, link performance analysis\cite{alfattani2021link}, and joint beamforming optimization\cite{wu2018intelligent}, etc. (see \cite{wu2021intelligent} for more references). 
The current literature focuses mainly on passive IRS due to its low power consumption and low hardware cost, which are important considerations for the practical deployment of IRS\cite{you2021wireless}. 
However, research findings demonstrate that the multiplicative fading of passive IRS poses a significant challenge for providing coverage enhancement to distant users. One option is to utilize a larger-scale IRS with more elements to enhance the performance gain, which yet increases the hardware cost and also the computational complexity for IRS phase optimization. An alternate solution is to exploit the recently proposed active IRS\cite{zhang2021active}, which helps alleviate the power loss due to multiplicative fading in IRS-reflected channel by incorporating power amplifiers to amplify/control the reflection amplitude in addition to phase.
In this paper, we focus on the deployment optimization of active/passive IRSs to enhance the communication coverage for a given set of users or Points of Interest (PoIs) in site-specific environments, and compare their difference in deployment strategies, key factors, and coverage performance. 

One of the key factors that is commonly ignored or simplified is the element radiation pattern (ERP)\cite{tang2022path} of IRS, as well as the antenna patterns of transmitters/receivers, which affects not only the link power budge, but also the fading statistics and hence ergodic rates. Our recent work in \cite{chen2023irs} proposes to deploy/integrate one passive IRS as part of a sectorized access point (AP), and reveals the significant impact of IRS ERP on the fading statistics and thus three-dimensional (3D) coverage performance. 
In this paper, we extend such modeling and consideration to both active and passive IRSs deployed distributively in the target area. Link-level performance evaluations confirm the non-negligible impact of IRS ERP as well as AP antenna pattern on the coverage performance.

\begin{figure}
	\centering
        \includegraphics[width=1\linewidth,  trim=80 120 130 20,clip]{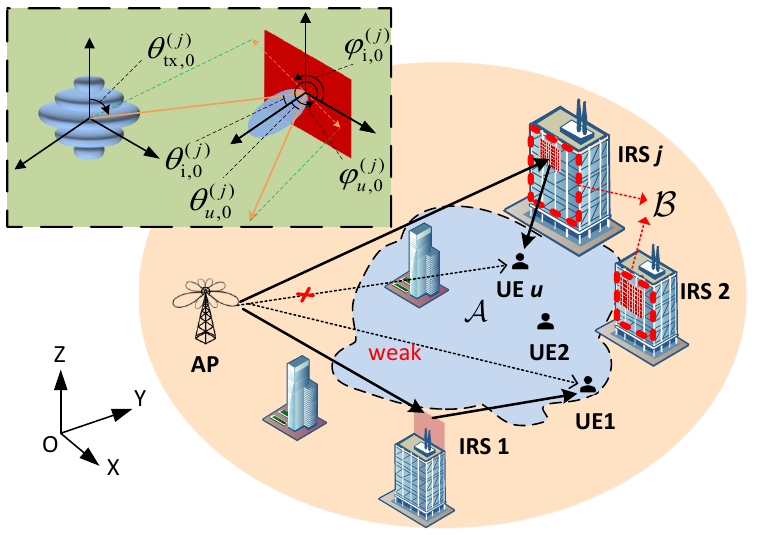}
	\caption{Site-specific deployment of IRSs for coverage enhancement.\vspace{-2ex}}\label{SystemModel}
\end{figure}

Another key factor that affects the IRS deployment decisions and associated coverage performance is the site-specific propagation environment. Existing research on IRS deployment typically assumes either blocked or non-blocked propagation paths between the AP, IRS and users without considering the specific physical source of blockages and the heterogeneous candidate sites for IRS deployment.
The literature on passive IRS, such as \cite{zhang2021intelligent}\cite{ling2021placement}, has investigated various deployment strategies without considering IRS ERP and site-specific environments. Similarly, studies on active IRS \cite{fu2022active}\cite{kang2023double} have explored IRS deployment issues and provided comparisons with passive IRS configurations, without considering IRS ERP and site-specific scenarios. Unlike these studies, our research undertakes a comprehensive analysis of both active and passive IRS deployment in site-specific, multi-building scenarios, taking into account the impact of IRS ERP and propagation environment on the deployment strategy.

    In this paper, we first investigate the link-level performance of active or passive IRS taking into account the IRS ERP as well as the antenna radiation pattern of the AP. Then the network-level coverage performance is evaluated/optimized in site-specific multi-building scenarios, by properly deploying multiple IRSs on candidate building facets to serve a given set of user equipments (UEs) or PoIs. The problem is reduced to an integer linear programming (ILP) based on given link-level metrics, which is then solved efficiently under moderate network sizes.
    Numerical results confirm the impact of AP antenna/IRS element pattern on the link-level performance. It is found that given the same total number of elements, passive IRS favors centralized deployment at a single spot in order to compensate for the multiplicative fading effect, while active IRSs could be deployed at multiple spots to improve both the network throughput and fairness. In addition, active IRSs, though associated with higher hardware complexity and cost, could significantly improve the site-specific network coverage performance in terms of average ergodic rate and fairness among the PoIs as well as the range of serving area, compared with passive IRSs that have a much larger number of elements.

\section{System Model}\label{SectionDesign}

\subsection{Positions of AP, IRSs, UEs, and Buildings}

As shown in Fig. \ref{SystemModel}, the AP is located on the Z-axis of the cartesian coordinate system, with a height of $ H_{\textrm{AP}}$. Distributed IRSs are installed on building surfaces to assist the AP in communicating with ground UEs, which are located in a target area  $\mathcal{A}$ with site-specific blockages. For simplicity and the purpose of illustration, buildings are modeled as cuboids with random heights. Each IRS is assumed to be a uniform planar array (UPA) with $N\triangleq N_{\textrm{h}} \times N_{\textrm{v}}$ elements. The spacing of the elements along the horizontal and vertical direction are $d_{\textrm{h}}$ and $d_{\textrm{v}}$, respectively, whose values are usually between $\frac{\lambda}{10}$ and $ \frac{\lambda}{2}$ \cite{tang2020wireless}, where $\lambda $ is the signal wavelength. The entire candidate area on the building surfaces for IRS deployment is denoted as $\mathcal{B}$. The actual deployment location of the IRS $j$ is denoted as $w_j \in \mathcal{B},j\in \mathcal{J}\triangleq\{1, \cdots,J\}$. 
%对于旋转角的表述先放着
% The rotation angle along the horizontal direction is denoted as $\Theta_{j}$, the rotation angle along the vertical direction is denoted as $\Phi_{j}$, the rotation angle is between $-\frac{\pi}{2}$ and $\frac{\pi}{2}$, and the positive and negative are determined according to the right-hand helix rule.

% The adjustable reflection coefficient of each reflection unit is defined as $\Gamma_{m,n}\triangleq A_{m,n} e^{j\phi_{m,n}}$. As a preliminary theoretical derivation, it is assumed that the reflection coefficient amplitude $A_{m,n}$ of all elements of IRS takes the same value, and its phases can be adjusted and aligned.

The coordinate of typical UE $u\in\mathcal{U}\triangleq\{1, \cdots,U\}$, in ground area $\mathcal{A}$ is expressed as $\mathbf{g}_u \triangleq \left[u_{\textrm{x}},u_{\textrm{y}},H_{\textrm{u}}\right]^T$. Taking the position of the central element as the IRS reference point, then the coordinate of the IRS $j$ reference coordinate is denoted by ${\mathbf{q}}_{j} \triangleq [x_j,y_j,z_j]^T$. 
\subsection{AP Antenna Pattern and IRS ERP}
\subsubsection{AP antenna pattern\cite{lyu2019network}}
 Assume that the AP antenna is composed of a uniform linear array (ULA) with several co-polarized dipole antenna elements placed vertically with equal spacing between elements. So the antenna pattern of the AP is a fixed pattern that is isotropic in the horizontal plane but vertically directional. In particular, the main beam of the fixed pattern is electrically down-tilted by $\theta_\textrm{tilt}$. 
 %关于fixed pattern的公式
\begin{comment}
 The power gain pattern of the ULA is given by
 \begin{equation}\label{BSantenna}
G_b(\theta)\triangleq G_e(\theta)\big(J(\theta)\big)^2=G_{e,\textrm{max}}\cos^2\theta \bigg(\frac{\sin(\frac{M}{2}\vartheta)}{\sqrt{M}\sin(\frac{1}{2}\vartheta)}\bigg)^2,
\end{equation}
 where $\theta\in(-90^\circ,90^\circ]$ is the elevation angle; $G_e(\theta)\triangleq G_{e,\textrm{max}}\cos^2\theta$ is the power gain pattern of each dipole antenna element with $G_{e,\textrm{max}}$ denoting its maximum value; and $J(\theta)\triangleq\frac{\sin(\frac{M}{2}\vartheta)}{\sqrt{M}\sin(\frac{1}{2}\vartheta)}$ is the array factor of the ULA with $\vartheta\triangleq \frac{2\pi}{\lambda}d_e(\sin\theta-\sin\theta_{\textrm{tilt}})$ in radian (rad). Note that in the downtilt direction, i.e., $\theta=\theta_{\textrm{tilt}}$, we have $\vartheta=0$ and $J(\theta_{\textrm{tilt}})=\sqrt{M}$, and hence  the antenna gain of AP ${G_\textrm{tx}}$ is
\begin{equation}\label{BSantennaTilt}
{G_\textrm{tx}}=G_b(\theta_{\textrm{tilt}})=M G_{e,\textrm{max}}\cos^2\theta_{\textrm{tilt}},
\end{equation}
further, the uniformed radiation pattern of AP can be expressed as ${F^{\textrm{tx}}}(\theta)=\frac{G_b(\theta)}{G_\textrm{tx}}$.
\end{comment}
The antenna gain and the uniform radiation pattern of the AP are denoted as ${G_\textrm{tx}}$ and ${F^{\textrm{tx}}}$, respectively. For a detailed expression of the fixed pattern, one can refer to the literature\cite{lyu2019network}.

\subsubsection{IRS ERP}
Consider the IRS with the actual ERP, denoted as $F^{\textrm{i}}{(\theta,\varphi )}$, and denote  $G_{\textrm{i}}$ as the maximum power gain of the element. The general expression of the ERP is as follows \cite{tang2022path}

\begin{equation}\label{IRSERPFormula}
F^\textrm{i}\left( {\theta ,\varphi } \right) = \left\{ {\begin{array}{*{20}{c}}
{(\cos \theta)^{G_\textrm{i}/2 - 1} ,}&{\theta  \in \left[ {0,\frac{\pi }{2}} \right],\varphi  \in \left[ {0,2\pi } \right],}\\
{0,}&{\theta  \in \left( {\frac{\pi }{2},\pi } \right],\varphi  \in \left[ {0,2\pi } \right].}
\end{array}} \right.
\end{equation}
where $\theta$ is the elevation angle and $\varphi$ is the azimuth angle. 
%关于Gi表达式
\begin{comment}
In general, $G_\textrm{i}$ can be expressed as
\begin{equation}\label{IRSERPFormula}
    G_\textrm{i}=\frac{4\pi }{\iint\limits_{\theta ,\varphi }{F_\textrm{i}(\theta ,\varphi )\sin \theta d\theta d\varphi }}.
\end{equation}
\end{comment}

%3D信道
\section{Site-Specific Channel Model and Impact of Radiation Pattern}
In this section, we introduce the distance/angle-dependent and site-specific channel model in the considered area, with emphasis on the impact of IRS ERP on the channel statistics.
Consider the downlink communication from a single AP to its served UEs or PoIs, denoted by $\mathcal{U} \triangleq \{1,\cdots, U\}$,
%a typical UE $u$ located at $\boldsymbol{w}_{u}$ and assigned with , 
whereas the results obtained can be similarly applied to the uplink communication as well. 
%Consider a single-cell downlink wireless communication system, where an IRS-aided sectorized BS is deployed to serve a set of $U$ UEs, denoted by $\mathcal{U} \triangleq \{1,\cdots, U \}$, which are uniformly distributed in a target area, $\mathcal{A}$. 
To focus on the coverage performance, assume that the served UEs are assigned with orthogonal-time Resource Blocks (RBs), i.e., time division multiple access (TDMA) or time-sharing is adopted.\footnote{TDMA is in general superior over FDMA due to hardware limitation of IRS passive reflection, which can be made time-selective, but not frequency-selective \cite{wu2021intelligent}. Other multiple access schemes are left for extended future work.}
For a typical UE $u\in\mathcal{U}$ assigned on a typical RB with bandwidth $B_\textrm{w}$, we introduce the IRS-related channels in the following.
Assume far-field propagation conditions for the AP-IRS, IRS-UE, and AP-UE link.
%, and all the channels are considered to be Rician fading channels.
% due to the presence of a large number of reflections and scatterers around it. 
% Assume for simplicity that the BSs and UEs are each equipped with a single antenna, while each IRS has $N$ reflecting elements.
The baseband equivalent channels from AP to IRS $j$, from IRS $j$ to UE $u$, and from AP to UE $u$ are denoted by $\mathbf {\tilde{h}}_{\textrm{i}}^{(j)}\triangleq[\tilde{h}_{\textrm{i},1}^{(j)},\cdots,{\tilde{h}}_{\textrm{i},N}^{(j)}]^T\in \mathbb{C}^{N\times 1}$, $\mathbf {\tilde{h}}_{\textrm{r},u}^{(j)}\triangleq[{\tilde{h}}_{\textrm{r},1,u}^{(j)},\cdots,{\tilde{h}}_{\textrm{r},N,u}^{(j)}]^T\in \mathbb{C}^{N\times 1}$, and ${\tilde{h}}_{\textrm{d},u}\in \mathbb{C}$, respectively.
% Let $\boldsymbol\phi^{(j)}\triangleq [\phi_1^{(j)}, \cdots,\phi_N^{(j)}]$ and 
Further, denote $\boldsymbol\Phi^{(j)}\triangleq \diag\{e^{\mathbf i\phi_1^{(j)}}, \cdots,e^{\mathbf i\phi_N^{(j)}}\}$ (with $\mathbf i$ denoting the imaginary unit) as the phase-shifting matrix of IRS $j$, where $\phi_n^{(j)}\in[0,2\pi)$ is the phase shift by element $n\in\mathcal{N}\triangleq\{1,\cdots,N\}$ of the IRS $j$ on the incident signal. $\boldsymbol P^{(j)}\triangleq \diag\{{ p_{1}^{(j)}}, \cdots,{ p_N^{(j)}}\}$ represents the amplification factor matrix of IRS $j$. In general, passive IRS element has amplification factor less than or equal to one.

%BS-IRS信道
\subsection{AP-IRS Channel}

When considering the isotropic radiation patterns on the AP and IRS sides, the channel amplitude from AP to the element $n$ of IRS $j$ can be expressed as\cite{chen2023irs}
\begin{equation}\label{gai}
|h_{\textrm{i},n}^{(j)}|\triangleq \sqrt{g_{\textrm{i}}^{(j)}}\xi_{\textrm{i},n}^{(j)},
\end{equation}
where $g_{\textrm{i}}^{(j)}$ denotes the average channel power gain, $\xi_{\textrm{i},n}^{(j)}$ accounts for channel fading and can be modeled as a random variable (RV) that characterizes the multi-path fading effect.
%xintong 关于新rician信道的表述
\begin{comment}
For simplicity, we consider the case with half-wavelength element spacing $d_\textrm{h}=d_\textrm{v}=\lambda/2$ and hence assume that the fading terms $\xi_{\textrm{i},n}^{(j)}$ of all IRS elements $n\in\mathcal{N}$ are independent and identically distributed (i.i.d.), which follow the Rician distribution $\textrm{Rice}(\upsilon,\sigma)$ with scale parameters $\upsilon = \sqrt{\frac{{K_{\textrm{i}}^{(j)}}}{{K_{\textrm{i}}^{(j)}}+1}}$ and $\sigma = \sqrt{\frac{1}{2({K_{\textrm{i}}^{(j)}}+1)}}$ and hence a mean power of ${\mathbb{E}\{\vert {\xi}_{\textrm{i},n}^{(j)}\vert ^2\}}=1$. In particular, the Rician factor ${K_{\textrm{i}}^{(j)}}$ represents the ratio of the mean power of the direct line-of-sight (LoS) path against that of all other non-LoS (NLoS) paths, which is distance- and/or angle-dependent in a given propagation environment, and also depends on the LoS/NLoS path-loss condition.
In the case of the NLoS path-loss condition, we assume ${K_{\textrm{i}}^{(j)}}=0$ which reduces to Rayleigh fading.
\end{comment}
For simplicity, we assume half-wavelength spacing between elements. Then the fading terms $\xi_{\textrm{i},n}^{(j)}$ can be considered as independent and identically distributed (i.i.d.), which follow the Rician distribution with a mean power of ${\mathbb{E}\{\vert {\xi}_{\textrm{i},n}^{(j)}\vert ^2\}}=1$ and the Rician K-factor ${K_{\textrm{i}}^{(j)}}$.

% In the case of LoS pathloss condition, we choose a distance/angle-dependent Rician K factor for ground/aerial UEs, respectively.
% For ground UEs, based on \cite{3GPPTR25996}, we have
% \begin{equation}
%     K=13-0.03 d_\textrm{bs}^{(j)}\quad (\textrm{dB}),
% \end{equation}
% which decreases with the BS-IRS $j$ distance $d_\textrm{bs}^{(j)}$ (m).

When the non-isotropic radiation patterns of AP and IRS are considered, the expression for channel amplitude becomes
\begin{equation}\label{gai}
|\tilde{h}_{\textrm{i},n}^{(j)}|\triangleq \sqrt{g_{\textrm{i}}^{(j)}}\tilde{\xi}_{\textrm{i},n}^{(j)}.
\end{equation}
The Rician factor of new channel fading becomes $\tilde{K}_{\textrm{i}}^{(j)}$, and the mean power becomes $\mathbb{E}\{\vert \tilde{\xi}_{\textrm{i},n}^{(j)}\vert ^2\}$. To describe the variation of small-scale fading, we respectively define the gain of the Rician factor $G_{K_\textrm{i}^{\textrm{(j)}}}$ and the mean fading power gain $\rho_\textrm{i}^{\textrm{(j)}}$ as
% defined to describe the change from $\xi$ to $\xi^{'}$, respectively.

% All the rice factors of new channel fadings become  $\tilde{K}_{\textrm{d},u}$, $\tilde{K}_{\textrm{i}}^{(j)}$, and $\tilde{K}_{\textrm{r},u}^{j}$, and powers becomes $\mathbb{E}\{\vert \tilde{\xi}_{\textrm{d},u}\vert ^2\}$, $\mathbb{E}\{\vert \tilde{\xi}_{\textrm{i},n}^{(j)}\vert ^2\}$ and $\mathbb{E}\{\vert \tilde{\xi}_{\textrm{r},n,u}^{(j)}\vert ^2\}$.

%重复太多了，三个信道都是rice fading，应该简洁表达

\begin{equation}
    G_{K_\textrm{i}^{\textrm{(j)}}} \triangleq \frac{\tilde{K}_{\textrm{i}}^{(j)}}{K_{\textrm{i}}^{(j)}},
\end{equation}
\begin{equation}
    \rho_\textrm{i}^{\textrm{(j)}} \triangleq \frac{
    \mathbb{E}\{\vert \tilde{\xi}_{\textrm{i},n}^{(j)}\vert ^2\}}
    {\mathbb{E}\{\vert {\xi}_{\textrm{i},n}^{(j)}\vert ^2\}}
    =\mathbb{E}\{\vert \tilde{\xi}_{\textrm{i},n}^{(j)}\vert ^2\}.
\end{equation}

% Define the rice factor gain as the ratio of the rice factor with a direction graph added to the rice factor without a direction graph added.
% That is
% \begin{equation}\label{IRSERPFormula}
%     {G_K} = \frac{{K'}}{K}
% \end{equation}
% The average fading power gain after adding the directional pattern is defined as $\rho$. That is
% \begin{equation}\label{IRSERPFormula}
% \rho  = \frac{{\mathbb{E}\{ {{{\left| {\xi '} \right|}^2}} \}}}{{\mathbb{E}\{ {{{\left| \xi  \right|}^2}} \}}} = \mathbb{E}\{ {{{\left| {\xi '} \right|}^2}} \}
% \end{equation}

Drawing upon reference\cite{chen2023irs} and \cite{yang2008impact}, we further derive the following expressions. The Rician factor gain and mean fading power gain can be respectively expressed as 
\begin{equation}\label{IRSERPFormula}
{G_{K_{\rm{i}}^{({\rm{j}})}}} = \frac{{{G_{{\rm{tx}}}}{G_{\rm{i}}}{F^{{\rm{tx}}}}(\theta _{{\rm{tx,0}}}^{(j)}){F^{\rm{i}}}(\Omega _{{\rm{i,0}}}^{(j)})}}{{(K_{\rm{i}}^{({\rm{j}})} + 1)E_{{\rm{i}},{\rm{NLoS}}}^{(j)}}},
\end{equation}
and
\begin{equation}
% \label{IRSERPFormula}
    \rho_\textrm{i}^{\textrm{(j)}} = \frac{{K_\textrm{i}^{\textrm{(j)}}}}{{{K_\textrm{i}^{\textrm{(j)}}} + 1}}{G_\textrm{tx}}{G_\textrm{i}}{F^\textrm{tx}}({\theta _{{\rm{tx,0}}}^{(j)}}){F^\textrm{i}}({\Omega _{{\rm{i,0}}}^{(j)}}) + {E_\textrm{i,NLoS}^{(j)}}.
\end{equation}
% \emph{Proof:} See Appendix \ref{Appen_pattern_RicianChannel}.\hfill $\blacksquare$
In the formulas, $E_\textrm{i,NLoS}^{(j)}$ is the mean power of the NLoS component when non-isotropic antennas are considered,
% ${G_\textrm{t}}$ and ${G_\textrm{r}}$ represent the gain of TX and RX, respectively.
${F^\textrm{tx}}({\theta _{{\rm{tx,0}}}^{(j)}})$ and ${F^\textrm{i}}({\Omega _{{\rm{i,0}}}^{(j)}})$ represent the values of TX and RX along the LoS direction, respectively. Here, ${\theta _{{\rm{tx,0}}}^{(j)}}$ denotes the elevation angle of the LoS direction of departure (DoD) for the AP, while ${\Omega _{{\rm{i,0}}}^{(j)}}\triangleq(\theta _{{\rm{i,0}}}^{(j)},\varphi _{{\rm{i,0}}}^{(j)})$ specifies the LoS direction of arrival (DoA) at the IRS.
%IRS-UE信道
\subsection{IRS-UE Channel}
Ground UEs are equipped with an isotropic antenna with unit antenna gain. Similarly, when considering the isotropic radiation patterns at both IRS and UE sides, the channel amplitude from the element $n$ of IRS $j$ to UE $u$ is given by
\begin{equation}\label{giu}
|h_{\textrm{r},n,u}^{(j)}|\triangleq \sqrt{g_{\textrm{r},u}^{(j)}}\xi_{\textrm{r},n,u}^{(j)},
\end{equation}
where, $g_{\textrm{r,u}}^{(j)}$ denotes the average channel power gains,  $\xi_{\textrm{r},n,u}^{(j)}$ account for channel fading. When the non-isotropic radiation pattern of  IRS is
considered, the expression for channel amplitude becomes
\begin{equation}\label{giu}
|\tilde{h}_{\textrm{r},n,u}^{(j)}|\triangleq \sqrt{g_{\textrm{r},u}^{(j)}}\tilde{\xi}_{\textrm{r},n,u}^{(j)},
\end{equation}
corresponding Rician factor gain $G_{K_{\textrm{r},u}^{\textrm{(j)}}}$ and average fading power gain $\rho_{\textrm{r},u}^{\textrm{(j)}}$ can be respectively expressed as 
\begin{equation}\label{IRSERPFormula}
    G_{K_{\textrm{r},u}^{\textrm{(j)}}} = \frac{{{G_\textrm{i}}{F^\textrm{i}}({\Omega _{u,0}^\textrm{(j)}})}}{({K_{\textrm{r},u}^{\textrm{(j)}}}+1){E_{\textrm{r},u,\textrm{NLoS}}^{(j)}}},
\end{equation}
and
\begin{equation}\label{IRSERPFormula}
    \rho_{\textrm{r},u}^{\textrm{(j)}} = \frac{{K_{\textrm{r},u}^{\textrm{(j)}}}}{{{K_{\textrm{r},u}^{\textrm{(j)}}} + 1}}{G_\textrm{i}}{F^\textrm{i}}({\Omega _{u,0}^{\textrm{(j)}}}) + {E_{\textrm{r},u,\textrm{NLoS}}^{(j)}},
\end{equation}
where, ${\Omega _{{{u,0}}}^{(j)}}\triangleq(\theta _{{{u,0}}}^{(j)},\varphi _{{{u,0}}}^{(j)})$ denotes the LoS DoD at the IRS.
%TX-IRS-UE级联信道
\subsection{AP-IRS-UE Channel}
Based on the above model and analysis, by considering non-isotropic radiation patterns at the AP and IRS, the cascaded AP-IRS $j$-UE $u$ channel can be written as

% \subsection{TX-IRS Channel}
% \begin{equation}\label{gai}
% |h_{\textrm{i},m,n}^{(j)}|^2\triangleq g_{\textrm{i},m}^{(j)}\xi_{\textrm{i},m,n}^{(j)}=\beta \big(r_{m,j}^2+(H_\textrm{B}-H_\textrm{I})^2\big)^{-\alpha/2}\xi_{\textrm{i},m,n}^{(j)},
% \end{equation}
\begin{equation}\label{hirmAll}
\mathbf {\tilde{h}}_{\textrm{ir},u}^{(j)}\triangleq [\mathbf {\tilde{h}}_{\textrm{r},u}^{(j)}]^T \boldsymbol P^{(j)}\Phi^{(j)} \mathbf {\tilde{h}}_{\textrm{i}}^{(j)}=\sum\limits_{n=1}^N {\tilde{h}}_{\textrm{i},n}^{(j)} {\tilde{h}}_{\textrm{r},n,u}^{(j)} p_{n}^{(j)}e^{\mathbf i\phi_n^{(j)}},
\end{equation}
where ${\tilde{h}}_{\textrm{ir},n,u}^{(j)}\triangleq {\tilde{h}}_{\textrm{i},n}^{(j)} {\tilde{h}}_{\textrm{r},n,u}^{(j)} p_{n}^{(j)} e^{\mathbf i\phi_n^{(j)}}$ denotes the AP-element $n$ of IRS $j$-UE $u$ channel.

% \subsubsection{Impact of IRS ERP on multipath fading statistics}
% \subsection{Impact of IRS pattern on multipath fading statistics}

%基站到用户的信道
\subsection{AP-UE Channel}
Similarly, when considering the isotropic radiation pattern, the channel amplitude from AP to UE $u$ is given by
\begin{equation}\label{gau}
|h_{\textrm{d},u}|\triangleq \sqrt{g_{\textrm{d},u}}\xi_{\textrm{d},u},
\end{equation}
where $g_{\textrm{d},u}$ is the average channel power gain, $\xi_{\textrm{d},u}$ accounts for channel small-scale fading. When the non-isotropic radiation pattern is considered, the above formula becomes
\begin{equation}
% \label{gau}
|\tilde{h}_{\textrm{d},u}|\triangleq \sqrt{g_{\textrm{d},u}}\tilde{\xi}_{\textrm{d},u}.
\end{equation}
The corresponding Rician factor gain $G_{K_{\textrm{d},u}}$ and average fading power gain $\rho_{\textrm{d},u}$ can be respectively expressed as 
\begin{equation}\label{IRSERPFormula}
    G_{K_{\textrm{d},u}} = \frac{{{G_\textrm{tx}}{F^\textrm{tx}}({\Omega _{u,0}^{\textrm{tx}}})}}{({K_{\textrm{d},u}}+1){E_{\textrm{d},u,\textrm{NLoS}}}},
\end{equation}
and
\begin{equation}
% \label{IRSERPFormula}
    \rho_{\textrm{d},u} = \frac{{K_{\textrm{d},u}}}{{{K_{\textrm{d},u}} + 1}}{G_\textrm{tx}}{F^\textrm{tx}}({\theta _{u,0}^{\textrm{tx}}}) + {E_{\textrm{d},u,\textrm{NLoS}}},
\end{equation}
where ${\theta _{{{u,0}}}^{\rm{tx}}}$ represents the elevation angle of LoS DoD from the AP towards the UE.

\subsection{Site-Specific LoS/NLoS Channel Model}
% Among them, all the channel fadings follow power normalized Rice distribution with the rice factor $K_{\textrm{d},u}$, $K_{\textrm{i}}^{(j)}$, and $K_{\textrm{r},u}^{j}$, respectively. 
All the aforementioned average channel power gains adopt the formulas for urban macro(UMa) scenario in 3GPP \cite{3GPP36777}. Due to site-specific blockages, the channels could be either LoS or NLoS. If the direct link between A and B is obstructed (A, B could refer to AP, IRS or UE), the average channel power gain is calculated by the NLoS path-loss model. Otherwise, it is calculated using the LoS path-loss model.
%使用los/nlos的公式
% \begin{equation}\label{LOS_NLOS}
% g_{\textrm{TR}}\triangleq
% \begin{cases}
%     g_{\textrm{LoS}},& \quad \textrm{without obstacles in between},\\
%     g_{\textrm{NLoS}}, & \quad \textrm{otherwise}.
% \end{cases}
% \end{equation}

% the Rician distribution $\textrm{Rice}(\upsilon,\sigma)$ with scale parameters $\upsilon = \sqrt{\frac{K}{K+1}}$ and $\sigma = \sqrt{\frac{1}{2(K+1)}}$ and hence a mean power of $\mathbb{E}[\vert \xi\rvert ^2]=1$.
 For the aforementioned Rician factors when considering the isotropic radiation patterns at both TX and RX sides, based on \cite{3GPPTR25996}, we have the following formula for ground UEs,
\begin{equation}
    K=13-0.03 d_{\textrm{TR}}\quad (\textrm{dB}),
\end{equation}
which decreases with the TX-RX distance $d_{\textrm{TR}}$ (m).

%接收功率和信噪比
\subsection{Received Power and SNR}
The signal received by UE $u$ can be expressed as \cite{zhang2021active}
\begin{equation}\label{IRSERPFormula}
        {r_{u}^{(j)}}  = \sqrt{p_u} ( {\tilde{h}_{\textrm{ir},u}^{(j)}}+{\tilde{h}_{\textrm{d},u}^{(j)}} )s_u + {\tilde{\mathbf{h}}_{\textrm{r},u}^{(j)}}\mathbf{P}^{(j)}\mathbf{\Phi}^{(j)} \mathbf{v} + n_u,
\end{equation}
where 
% $\textrm{f}$ $\textrm{P}$ $\Theta$ and $\textrm{g}$ represent the channel matrix between IRS and UE, the amplification factor matrix of IRS, the phase matrix of IRS, and the channel matrix between BS and IRS, respectively. 
$p_u$ represents the power sent by AP to UE $u$, $s_u$ represents the signal sent to UE $u$ by AP, and satisfy $\mathbb{E}\{|s_u|^2\}  = 1$, $\mathbf{v}\in \mathbb{C}^{N\times 1}$ represents the dynamic noise inside the active IRS, which is set to 0 for passive IRS. $n_u$ represents the thermal noise of the UE receiver.

The received SNR of UE $u$ aided by IRS $j$ can be expressed as
\begin{equation}\label{IRS_SNR_prima}
{\eta_{u}^{(j)}} = \frac{{{p_u}{{ {\left| {{\tilde{h}_{\textrm{ir},u}^{(j)}}+{\tilde{h}_{\textrm{d},u}}} \right|} }^2}}}{\Vert{\tilde{\mathbf{h}}_{\textrm{r},u}^{(j)}}\mathbf{P}^{(j)}\mathbf{\Phi}^{(j)}\Vert^{2}\sigma_v^{2} + {\sigma ^{2}}},
\end{equation}
where ${\sigma ^{2}} \triangleq N_0 B_{\rm{w}}$ represents the noise power of additive white Gaussian noise (AWGN) at UE $u$. Here, $N_0$ and $B_{\rm{w}}$ represent the power spectral density (PSD) of the noise and bandwidth, respectively. Similarly, $\sigma_v^{2} \triangleq {N}_v B_{\rm{w}}$ represents the active noise power of each element. Here, $N_v$ represents PSD of the active noise at each element. 
Assume that each active IRS element has the same amplification factor (i.e., $p_n^{(j)} := p^{(j)}$ for all $n \in \mathcal{N}$). Define that $P_\textrm{A}^{(j),\textrm{max}}$ represents the maximum power consumed on active IRS $j$, and $P_u^{\textrm{max}}$ represents the maximum power transmitted by AP to UE $u$. Assume that each active IRS has the same maximum reflect power (i.e., $P_\textrm{A}^{(j),\textrm{max}} := P_\textrm{A}^{\textrm{max}}$ for all $j \in \mathcal{J}$). Assume that perfect channel state information (CSI) can be obtained through channel estimation methods, then the optimal SNR for UE $u$ is achieved when the following conditions are satisfied\cite{zhang2021active}:
% When the power $p_u$ is transmitted by the BS to UE $u$, the phase shift $\phi_n$ of each reflection unit of the active IRS and the amplification factor $p^{(j)}$ are set as follows, The SNR of the user reached the maximum value.
\begin{subequations}\label{sub_optimization_result}
\begin{equation}\label{IRSERPFormula}
    {p_u^{\textrm{opt}}} = P_{u}^{\textrm{max}},
\end{equation}
\begin{equation}\label{IRSERPFormula}
    {\phi _n^{(j),\textrm{opt}}} =  - \angle ( \tilde{h}_{\textrm{i},n}^{(j)},{\tilde{h}_{\textrm{r},n,u}^{(j}} ) + \angle \tilde{h}_{\textrm{d},u},\forall n \in \left\{ {1, \cdots ,N} \right\},
\end{equation}
\begin{equation}\label{IRSERPFormula}
    {p^{(j),\textrm{opt}}} = \sqrt {
    \frac{
    P_{\textrm{A}}^{(j),\textrm{max}}}
    {{p_u^{\textrm{max}}}{\mathop \sum \nolimits_{n = 1}^N {{\vert {{\tilde{h}_{\textrm{i},n}^{(j)}}} \vert}^2 + N\sigma_v^{2}}}
    }},\forall n \in \left\{ {1, \cdots ,N} \right\}.
\end{equation}
    % {P_\textrm{u}^{\textrm{max}} \mathop \sum \nolimits_{n = 1}^N {{{ | {h_{\textrm{i},n}^{(j)}} | }^{2}} + N \sigma_v^{2}}} 
    % {{P_\textrm{u}^{\textrm{max}} \sum_{n = 1}^N |h_{\textrm{i},n|}^{(j)} + N \sigma_v^2}}}
\end{subequations}

% }
%假设相位对齐和幅度一致后的简化信噪比公式
By substituting (\ref{sub_optimization_result}) into (\ref{IRS_SNR_prima}), the optimal SNR can be obtained as
\begin{equation}\label{IRS_SNR_simple_P_phase}
\begin{aligned}
{\gamma_{u}^{(j)}} &=
\frac{
{{P_u^{\textrm{max}}}{{{\left| {p^{(j),\textrm{opt}}} \mathop \sum \nolimits_{n = 1}^N {{\vert {{\tilde{h}_{\textrm{i},n}^{(j)}}} \vert}{\vert {{\tilde{h}_{\textrm{r},n,u}^{(j)}}} \vert}} + \vert{\tilde{h}_{\textrm{d},u}}\vert \right|} }^2}}
}
{
({p^{(j),\textrm{opt}}})^{2}\sigma_v^{2}{\mathop \sum \nolimits_{n = 1}^N {{\vert {{\tilde{h}_{\textrm{r},n,u}^{(j)}}} \vert}^2}} + {\sigma ^{2}}
}\\&=
\frac{
{{P_u^{\textrm{max}}}{{{\left| \sqrt{P_\textrm{A}^{(j),\textrm{max}}} \mathop \sum \nolimits_{n = 1}^N {{\vert {{\tilde{h}_{\textrm{i},n}^{(j)}}} \vert}{\vert {{\tilde{h}_{\textrm{r},n,u}^{(j)}}} \vert}} + \sqrt{T}\vert{\tilde{h}_{\textrm{d},u}}\vert \right|} }^2}}
}
{
{P_\textrm{A}^{(j),\textrm{max}}}\sigma_v^{2}{\mathop \sum \nolimits_{n = 1}^N {{\vert {{\tilde{h}_{\textrm{r},n,u}^{(j)}}} \vert}^2}} + {\sigma ^{2}}T
},
\end{aligned}
\end{equation}
where, $T={P_u^{\textrm{max}}}{\mathop \sum \nolimits_{n = 1}^N {{\vert {{\tilde{h}_{\textrm{i},n}^{(j)}}} \vert}^2 + N\sigma_v^{2}}}$.

\subsection{Ergodic Throughput and Coverage Indicator}

The corresponding instantaneous rate of UE $u$ aided by IRS $j$ is expressed in bps/Hz as
\begin{equation}
    r_u^{(j)} = \log (1 + \gamma _u^{(j)}),
\end{equation}
and the ergodic throughput can be expressed as 
\begin{equation}
     % \tilde{R}_u^{(j)} = \mathbb{E} \{ r_u^{(j)} \}.
    {R}_u^{(j)} = \mathbb{E} \{ r_u^{(j)} \}.
\end{equation}
\begin{comment}
Define the ergodic throughput  of UE $u$ when the best IRS is selected to serve it as
\begin{equation}
{R}_u = \max_{j \in \mathcal{J}} R_u^{(j)}.
\end{equation}
The average ergodic throughput of all UEs can be expressed as
\begin{equation}
\bar{R} = \frac{1}{|\mathcal{U}|}  {\textstyle \sum_{u\in\mathcal{U}}} R_{u}.
\end{equation}
\end{comment}
In addition, the average SNR in dB can be expressed as 
\begin{equation}
    {\bar{\gamma}_{u}^{(j)}} = 10\cdot \log_{10}{ \mathbb{E} \{ \gamma_{u}^{(j)} \} }.
\end{equation}
Denote $\overline{\rm{SNR}}$ (dB) as the required average SNR level. Then the coverage indicator for UE $u$ aided by IRS $j$ can be defined as
\begin{equation}
    C_u^{(j)} =
    \begin{cases}
        1, & \text{if ${\bar{\gamma}_{u}^{(j)}} \ge \overline{\rm{SNR}}$}, \\
        0, & \text{otherwise}.
    \end{cases}
\end{equation}

Finally, note that due to the complexity of the SNR expression in \eqref{IRS_SNR_simple_P_phase}, closed-form expressions for the ergodic rate and coverage indicator can not be found. Therefore, we resort to Monte Carlo (MC) simulations to evaluate them by averaging over a large corpus of channel fading samples. Fortunately, the per-link ergodic throughput or coverage indicator needs to be calculated only once for a given UE served by a given IRS, and the computational time is negligible compared with the time it takes to make practical deployment decisions.

\begin{comment}
    
Noting that $\bar R$ is the instantaneous rate threshold, the non-outage probability $P_{{\rm{no}},u}^{(j)}$ for IRS $j$ auxiliary UE $u$ is expressed as follows
\begin{equation}
    {\rm{P}}_{{\rm{no}},u}^{(j)} \buildrel \Delta \over 
= \left\{ {{{\log }_2}\left( {1 + \gamma _u^{(j)}} \right) \ge \bar R} \right\},
\end{equation}
then the throughput of UE $u$ with the assistance of IRS $j$ can be obtained as
\begin{equation}
    v_u^{(j)} = {\rm{P}}_{{\rm{no}},u}^{(j)}\bar R
\end{equation}
\end{comment}

\section{Problem Formulation and Optimization}\label{SectionOptimization}
%非中断概率的定义先不要
% Define the non-outrage probability of UE communication, that is, the probability of receiving a signal, expressed as $P_{non}$
% \begin{equation}\label{IRSERPFormula}
%     {P_{{\rm{non}}}} \buildrel \Delta \over = {\rm{P}}\left\{ {\rho ' \ge {\rho _0}} \right\},
% \end{equation}
% In the formula, $\rho_0$ is the SNR threshold for signals received by the UE. $\bar{R}$ is defined as the minimum communication rate that UE needs to meet. Then the throughput of UE,  $\upsilon$ be expressed as
% \begin{equation}\label{IRSERPFormula}
%     \upsilon  \buildrel \Delta \over = {P_{{\rm{non}}}}\bar R.
% \end{equation}
% For the IRS-UE channel, it may be a line-of-sight channel or a non-line-of-sight channel. Here, whether it is a line-of-sight channel is judged by whether the actual obstacle is blocked or not. Considering IRS candidate deployment region $B_1$,$\ldots$,$B_J$, the problem of multiple IRSs field deployment to maximize the minimum throughput of users in target region $\mathcal{A}$ is studied, which can be expressed as follows:
Our research focuses on system-level performance optimization, and aims to maximize the average ergodic throughput of all the UEs in the region $\mathcal{A}$ by jointly optimizing the IRS locations and the association between the UEs and IRSs. The problem can be formulated as follows:
\begin{align}\label{P0}
    & (\text{P0}): \max_{{w_j},j \in \mathcal{J}}  \frac{1}{|\mathcal{U}|}  {\textstyle \sum_{u\in\mathcal{U}}} R_{u} \\
    & \text{s.t.} \quad {R}_u = \max_{j \in \mathcal{J}} R_u^{(j)},
    \label{P0_c1} \\
    &\phantom{\text{s.t.}} \quad {w_j} \in {\mathcal{B}},\forall j \in \mathcal{J}.\label{P0_c2} 
\end{align}
%阐述原问题的求解难度，引出离散化表述
% the typical user $u$ given the IRS deployment situation, chooses the IRS that makes the largest SNR. Note that as a preliminary exploration, only one IRS is considered for each candidate region, and only one IRS is deployed for each user.
The constraint (\ref{P0_c1}) represents the maximum ergodic throughput of UE $u$ with optimal IRS association. In addition to the combinatorial complexity of UE-to-IRS association, another main challenge lies in the continuous (infinite) candidate deployment locations of the IRS on the building surface. Due to site-specific blockages, the channel variations across continuous IRS locations could be drastically different due to possible changes of the LoS/NLoS propagation condition. As a result, there is no closed-form continuous expression for the performance metric (e.g., ergodic throughput) as a function of IRS candidate locations, and hence traditional continuous optimization methods (e.g., convex optimization) are not applicable. 

To this end, we propose to discretize the candidate IRS deployment sites into grid spots and reformulate the problem into an integer linear programming (ILP) based on given per-link ergodic throughput $R_u^{(j)}$. The building surface can be divided into grids with a proper resolution, thereby reducing the number of candidate IRS locations. 
Further reduction can be achieved by applying filtering strategy based on the LoS/NLoS status of AP-IRS links, since the LoS condition is prefered in general. By applying such filtering, the candidate IRS deployment spots could be irregularly distributed in the target region, making up a set $\mathcal{M}=\{1,\cdots,M\}$.

As a preliminary study, we consider deploying at most one IRS per candidate grid point, and each UE is served by only one IRS, with no upper limit on the number of UEs served by each IRS. The problem can be reformulated as
\begin{align}\label{P1}
    &\text{(P1)}: \max_{a_{u,j}, b_m, \forall u,j}  \frac{1}{|\mathcal{U}|}\sum_{u=1}^{U} \sum_{j=1}^{J} a_{u,j} {R}_u^{(j)} \\
    &\text{s.t.} \, \sum_{j=1}^{J} a_{u,j} = 1, \forall u\in\mathcal{U};  \label{PR0_c1} \\
    &\phantom{\text{s.t.}} \, \sum_{m=1}^{M} b_m = J; \label{PR0_c2} \\
    &\phantom{\text{s.t.}} \, a_{u,j} \le b_m, \forall u\in\mathcal{U},  j\in\mathcal{J}, m\in\mathcal{M};  \label{PR0_c3} \\
    &\phantom{\text{s.t.}} \, a_{u,j}, b_m \in \{0,1\}, \forall u\in\mathcal{U}, j\in\mathcal{J}, m\in\mathcal{M}.\label{PR0_c4}
\end{align}
Here ${a_{u,j}}$ is a binary variable indicating the association between IRS $j$ and UE $u$, which is set to 1 if IRS $j$ serves UE $u$, or 0 otherwise.. Similarly, for each candidate deployment grid point $m \in \mathcal{M}$, ${b_{m}} $ denotes a binary variable which equals to 1 when an IRS is deployed at candidate deployment grid point $m$, or 0 otherwise.

Although the complexity has been reduced, problem (P1) is still NP-hard in determining the optimal IRS deployment. When considering $M$ candidate deployment grid points and $J$ IRSs, the total number of possible combinations is given by $C_M^J = \frac{{M!}}{{J!(M - J)!}}$. Classic solution methods include heuristics, metaheuristics, branch-and-bound, etc.. 
% Ordinary heuristics, such as tabu search, greedy algorithms, and exchange algorithms, can sometimes provide a better feasible solution quickly, but the quality of the solution is not guaranteed. Exact class algorithms, such as branch-and-bound, often grapple with the issue of dimension explosion. Typically, the solution time increases exponentially with the problem size. 
Fortunately, for our considered single-AP scenario serving moderately large target regions, we could leverage on state-of-the-art commercial solver such as Gurobi\cite{gurobi10} (version 10.0.1) to solve it, with default parameter settings. More explorations on the algorithm design to be applied for larger network sizes are left for our future work.

\section{Numerical Results}
% the carrier frequency $f_c$=2GHz, the bandwidth $B_w$=200KHz, the noise power of active IRS $N_v$=160dBm/Hz, the thermal noise of receiver $N_0$=-174dBm/Hz, and the number of IRS elements $N$=64
This section provides numerical evaluations on both the link-level performance and the site-specific network performance.
Each MC result is based on the average of 10000 random fading samples. 
% , as well as performance comparison analysis of optimization algorithms.
The following parameters are used if not mentioned otherwise\cite{zhang2021active}\cite{chen2023irs}: the carrier frequency $f_{\rm{c}}=$ 2 GHz, $B_{\rm{w}}=$ 200 KHz, $N_v=$ -160 dBm/Hz, $N_0=$ -174 dBm/Hz, $H_{\rm{AP}}=$ 25 m, $H_{{u}}=$ 1.5 m, $G_{\rm{i}}=$ 6 dBi (i.e. $\cos{\theta}$), the total power $P_{\rm{total}}=$ 10 mW, $P_{u}^{\textrm{max}}=5$ mW, $P_{\rm{A}}^{\textrm{max}}=$ 5 mW. When no IRS is deployed, the transmit power at the AP is equal to $P_{\rm{total}}$.

\begin{figure}
	\centering
        \includegraphics[width=1\linewidth, trim=40 60 120 85,clip]{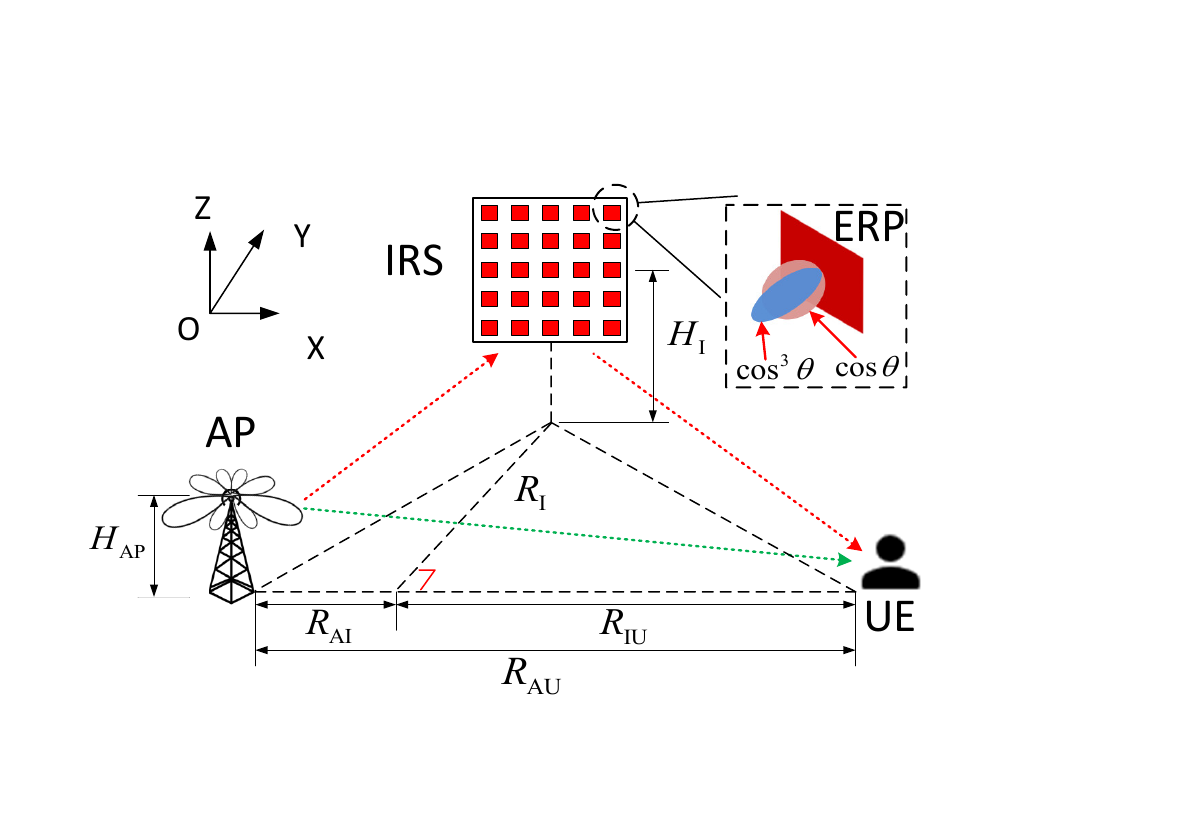}
        \caption{Single IRS deployment link performance experiments.\vspace{-2ex}}
        \label{link_model_fig}
\end{figure}

\subsection{Link Performance Analysis}
\begin{figure}
	\centering
        \includegraphics[width=1\linewidth, trim=60 0 60 30,clip]{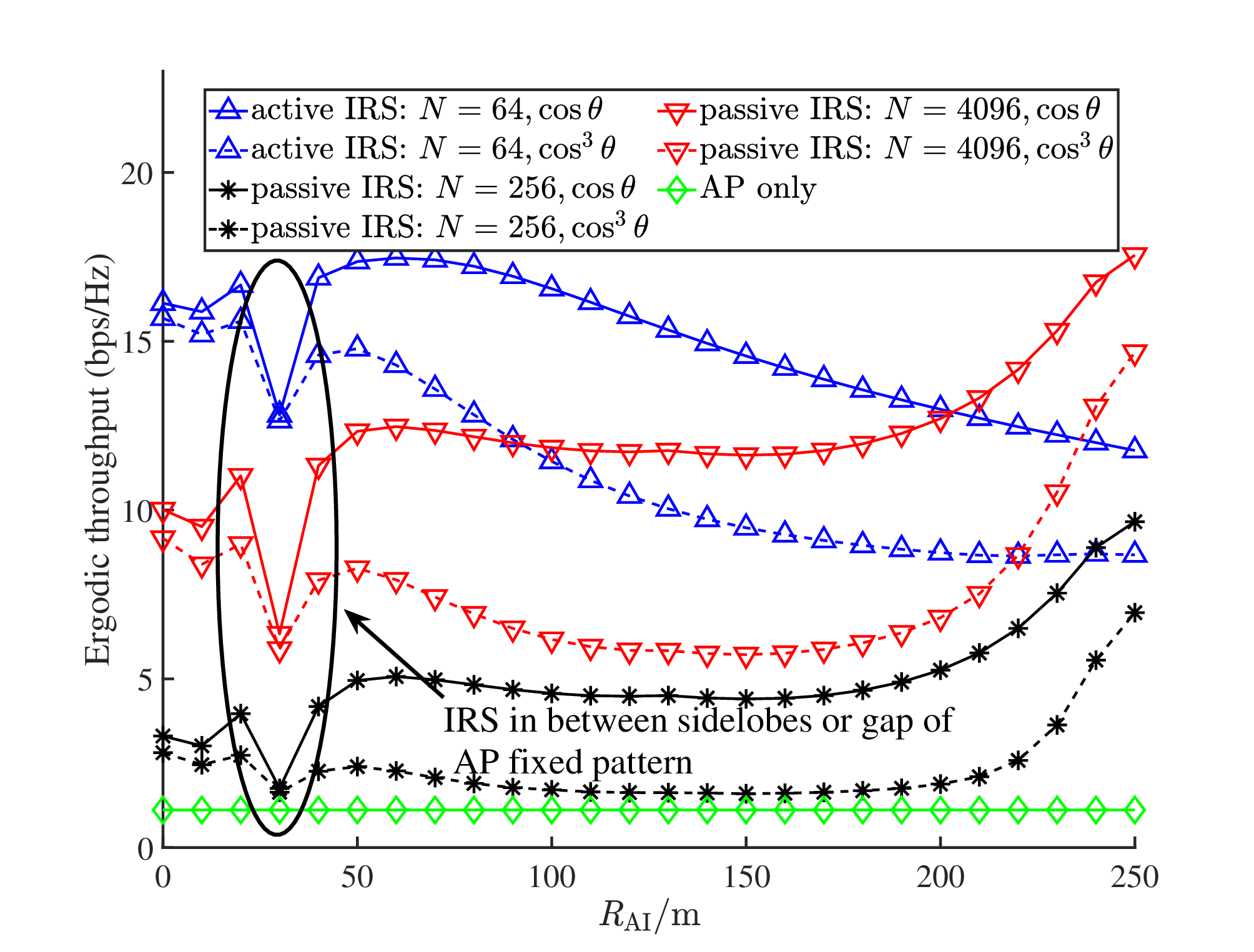}
        \caption{Ergodic throughput of the UE aided by IRS or AP-only.\vspace{-2ex}}
        \label{fig:pno_MC_theory_link}
\end{figure}

In order to analyze the link performance, we consider the scenario depicted in Fig. \ref{link_model_fig}, with additional parameters given by $R_{\rm{AU}}=250$ m, $R_{\rm{I}}=10$ m and $H_{\rm{I}}=10$ m. 
% \begin{figure}
% 	\centering
% 	%\includegraphics[width=0.8\linewidth,  trim=0 30 0 20,clip]{model_bf.eps}
%         \includegraphics[width=0.8\linewidth]{model_bf.eps}
% 	\caption{model\_bf.\vspace{-2ex}}
%         \label{model_bf}
% \end{figure}
% \begin{figure}
%         \centering
%         \begin{subfigure}{0.24\textwidth}
%     	\centering
%     	%\includegraphics[width=0.8\linewidth,  trim=0 30 0 20,clip]{model_bf.eps}
%             \includegraphics[width=1\linewidth, trim=0 0 0 0,clip]{link_simulation_1.eps}
%     	\caption{model\_bf.\vspace{-2ex}}
%             \label{model_bf}        
%         \end{subfigure}
%         \begin{subfigure}{0.24\textwidth}
% 	    \centering
%     	%\includegraphics[width=0.8\linewidth,  trim=0 30 0 20,clip]{model_bf.eps}
%             \includegraphics[width=1\linewidth, trim=20 0 30 0,clip]{simple_link_scene.png}
%     	\caption{simple link scene.\vspace{-2ex}}
%             \label{fig:simple_link_scene}         
%         \end{subfigure}
%         \caption{two\_simple\_link}
%         \label{fig:two_simple_link}
% \end{figure}
% Here the antennas for AP and UE are assumed to be isotropic. Active and passive IRS are considered separately. The element radiation pattern is modeled as $\cos\theta$ and the half-space isotropic pattern (i.e., hemisphere with directional gain = 2).
The signals received by the UE have two types: direct signals from the AP and reflected signals from the IRS. We assum LoS links for AP-IRS and IRS-UE, and the NLoS link for AP-UE. Different ergodic throughput for the UE can be achieved when the IRS's horizontal distance $R_{\rm{AI}}$ is changed. To draw a comparison, experiments were conducted with different number of IRS elements. Moreover, two ERPs for the IRS are considered, with $G_{\rm{i}}=$ 6 dBi (i.e. $\cos{\theta}$) and $G_{\rm{i}}=$ 9 dBi (i.e. $\cos^3{\theta}$), respectively. We also consider a baseline case without IRS.
% Rice channel and uniform scattering models are taken into account for all channels. (Rice factor what model? -second opening code). For fairness, the same total power is considered for passive and active IRS-assisted link communication. In the passive IRS-assisted scenario, 10mW of AP power is considered, whereas in the active IRS-assisted situation, 5mW of AP power is considered and 5mW of power is allocated to the active IRS. 
The simulation results are shown in Fig. \ref{fig:pno_MC_theory_link}.

% \begin{figure}
%         \centering

%         \begin{subfigure}{0.24\textwidth}
%     	\centering
%     	%\includegraphics[width=0.8\linewidth,  trim=0 30 0 20,clip]{model_bf.eps}
%             % \includegraphics[width=0.9\linewidth, trim=0 0 0 0,clip]{link_active_simulation.png}
%             \includegraphics[width=0.9\linewidth, trim=0 0 0 0,clip]{pno_MC_theory_link.png}
%     	\caption{model\_bf.\vspace{-2ex}}
%             \label{link_active_simulation}            
%         \end{subfigure}
%         \begin{subfigure}{0.24\textwidth}
% 	    \centering
%     	%\includegraphics[width=0.8\linewidth,  trim=0 30 0 20,clip]{model_bf.eps}
%             % \includegraphics[width=0.9\linewidth, trim=0 0 0 0,clip]{link_passive_simulation.png}
%             \includegraphics[width=0.9\linewidth, trim=0 0 0 0,clip]{averageRate_MC_link.png}
%     	\caption{model\_bf.\vspace{-2ex}}
%             \label{link_passive_simulation}            
%         \end{subfigure}

%         \caption{passive\_active\_simulation}
%         \label{fig:passive_active_simulation}

% \end{figure}

It can be seen that deploying IRS (passive or active) brings significant performance improvement. Notice that when the $R_{\rm{AI}}=50$ m, the active IRS ($N=64$) outperforms the passive IRS ($N=256$) significantly, with the gap up to about 12bps/Hz. However, on the UE side, the performance of the active IRS ($N=64$) and the passive IRS ($N=256$) are close to each other. This is primarily due to the increased effect of active noise when the active IRS is close to the UE. 
Additionally, it has been observed that when the total number of passive IRS elements is increased, such as $N=4096$, its link performance improves significantly and can surpass the active IRS at some locations. Nonetheless, it should be noted that in practical implementations, a passive IRS with a large number of elements may lead to increased cost as well as challenges in channel estimation and beamforming. 
Moreover, the overall performance corresponding to the ERP of $\cos{\theta}$ is better than that of the ERP of $\cos^3{\theta}$. This is because the more focused antenna gain leads to diminished gain when the incident and departure angles deviate from the center of the beam.
Furthermore, the fixed pattern of the AP impacts the link performance. When the signal falls into the sidelobes or gap of the AP's fixed pattern, a noticeable degradation in performance can occur.

\begin{figure}
	\centering
        \includegraphics[width=1\linewidth,  trim=60 30 40 30,clip]{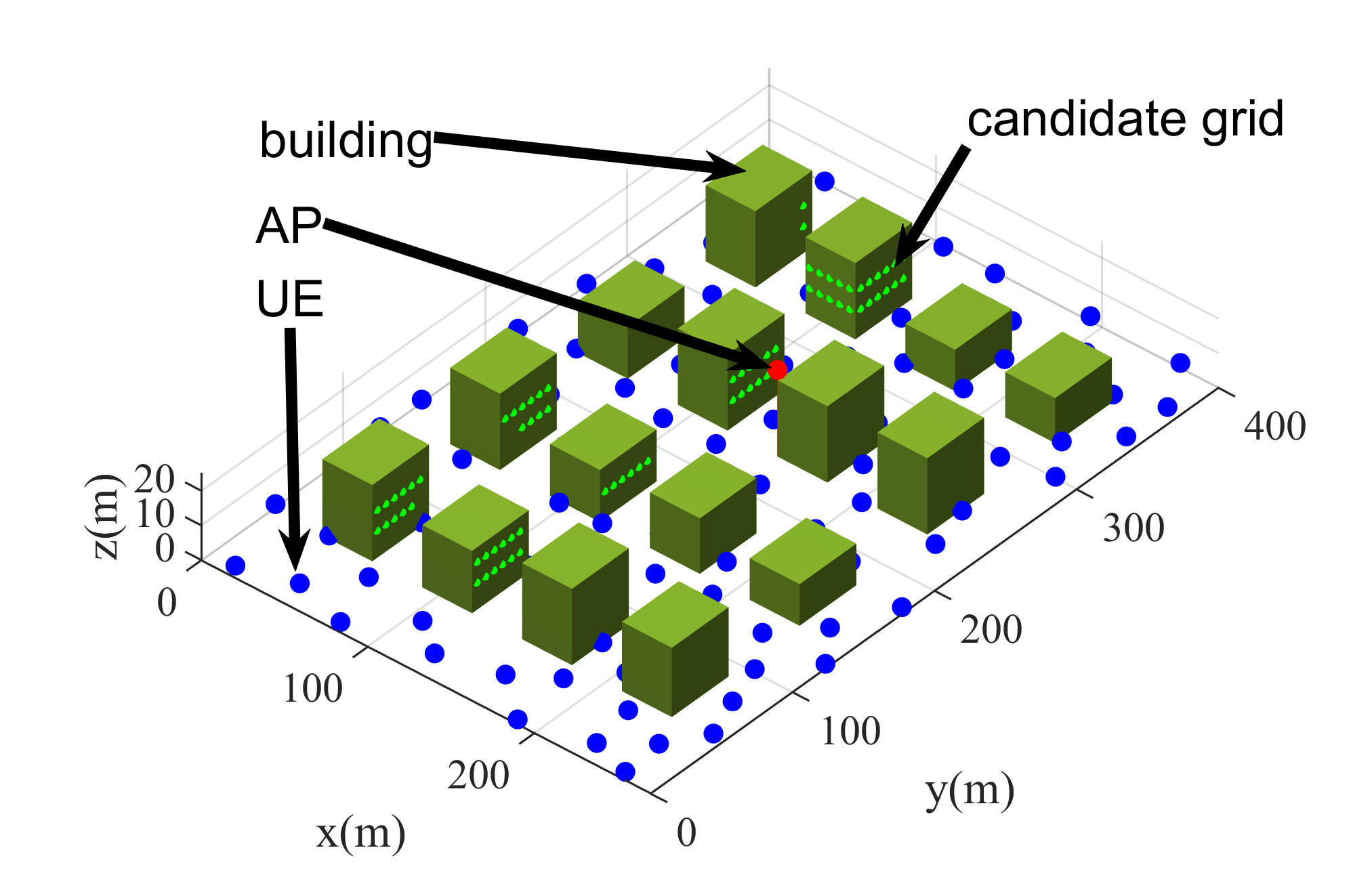}
	\caption{Site-specific deployment of IRS to enhance coverage, considering the grids as candidate deployment locations, 100 UEs and 149 candidate deployment locations.\vspace{-2ex}}
        \label{median_scene}
\end{figure}

\subsection{IRS Deployment Given Total Number of Elements}
As shown in Fig. \ref{median_scene}, we consider a scenario with $4\times4=16$ buildings within an area of 270 m by 400 m. Each building has a length of 30 m and a width of 40 m. The building heights are randomly generated within a range of 12 to 22 meters. Additionally, 100 UEs or PoIs are randomly distributed on the streets, and there are 149 grid points available for IRS deployment. 

\begin{figure}
    \centering
        \includegraphics[width=1\linewidth, trim=45 0 0 30,clip]{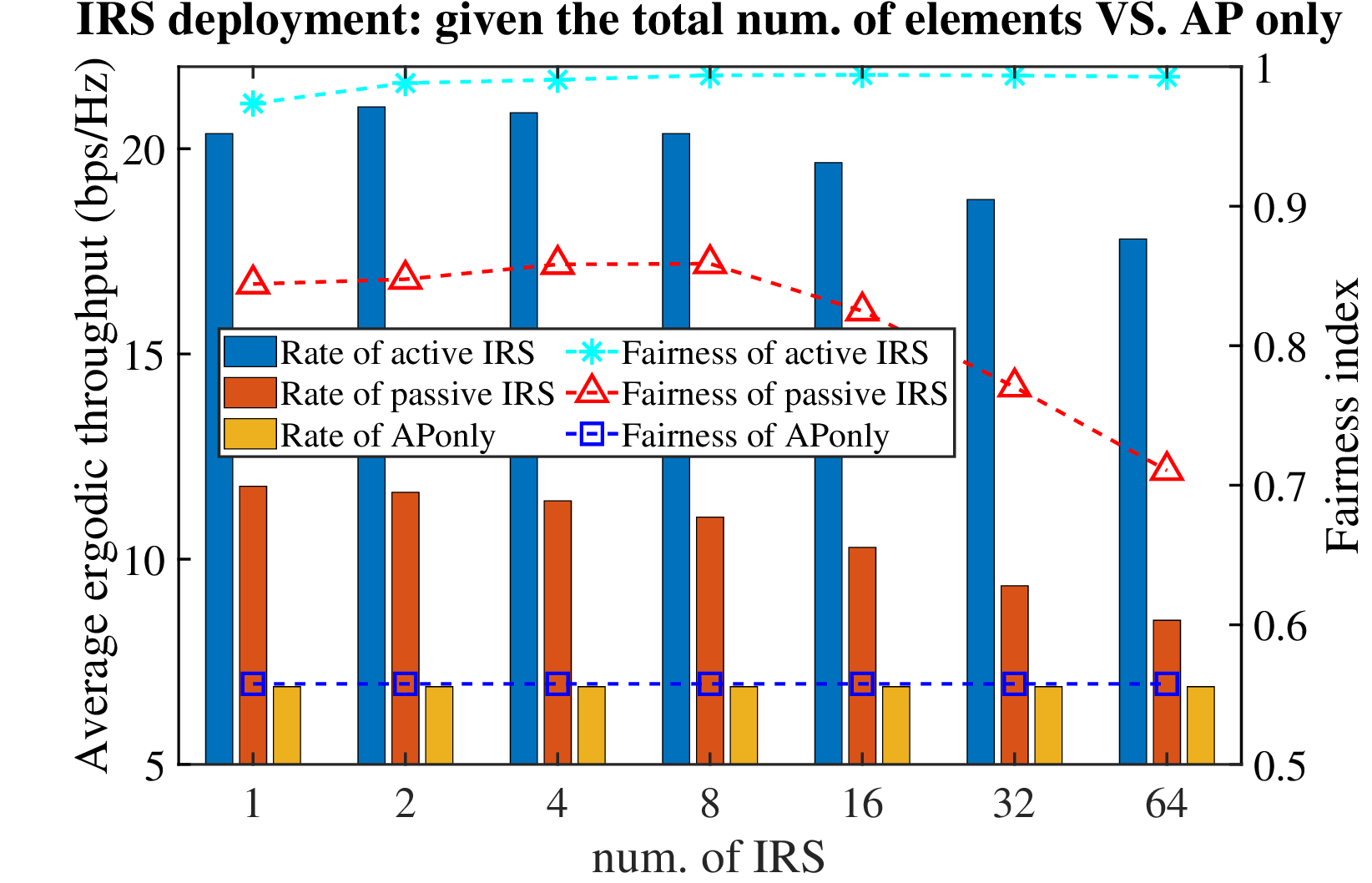}
        % \caption{Comparison experiment of active IRS, passive IRS, and no IRS deployment.\vspace{-2ex}}
        \caption{IRS deployment given total number of elements $N=1024$.\vspace{-2ex}}
        \label{IRS_deployment-given_total_number}
\end{figure}

This section discusses the performance variation when a total number of elements are equally divided into different number of distributed IRSs. The experimental setup is similar to the previous section, with the same total number of elements $N=1024$ for both active and passive IRS.
The results are shown in Fig. \ref{IRS_deployment-given_total_number}.
It is found that to maximize average ergodic throughput, there exists a more suitable division for the active IRS. For example, in the considered setup here, it's best to divide into 2 active IRSs. On the other hand, for the passive IRS, dividing it into more IRSs does not bring the expected benefits. This is because the more divisions there are, the smaller each IRS becomes, and the benefits that each UE receives from a single IRS become much smaller as well.
Notice that, in terms of the fairness index\footnote{This index is defined as $J\triangleq \frac{(\sum_{u\in\mathcal{U}} R_u)^2}{|\mathcal{U}| \sum_{u\in\mathcal{U}} R_u^2}$, where $J\in(0,1]$ and a higher $J$ represents better fairness.}, 
% increasing the number of divisions enhances the fairness among UEs for active IRS, whereas it decreases fairness for passive IRS when the number of divisions is too much (number of passive IRS is bigger than 8).
the fairness of active IRS is improved with more divisions, but the fairness of passive IRS is decreased with more divisions. This is because passive IRS is more suitable for deploying at the AP side or the UE side. When passive IRS is deployed at the UE side, it only serves the UEs in its vicinity, which results in a wider gap in the performance among the UEs.

\begin{figure}
    \centering
        \includegraphics[width=1\linewidth, trim=10 110 50 30,clip]{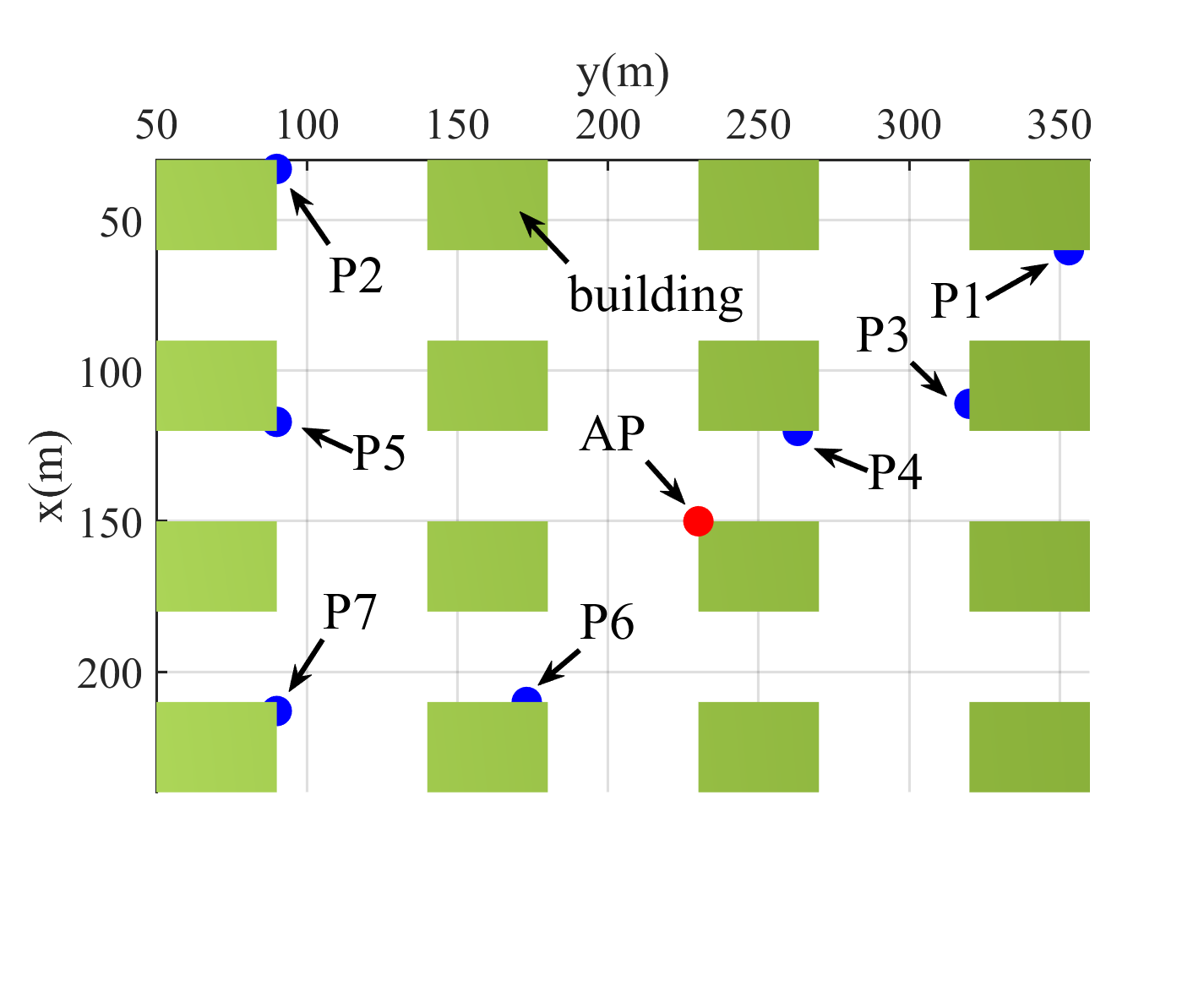}
        % \caption{Comparison experiment of active IRS, passive IRS, and no IRS deployment.\vspace{-2ex}}
        \caption{Optimized IRS deployment positions. For active IRSs, when the total elements are divided into 1, 2, and 4 IRSs, the optimal deployment locations are (P4), (P4, P7), and (P1, P2, P4, P7), respectively. For passive IRSs, the optimal deployment locations under the same dividing schemes are (P4), (P4, P5), (P3, P4, P5, P6), respectively.\vspace{-2ex}}
        \label{position_deploy_1}
\end{figure}

%新增对最优部署位置的解释
To visualize the optimal deployment results, the dividing schemes with 1, 2, and 4 IRSs are plotted as shown in Fig. \ref{position_deploy_1}. The result indicates that when the total elements are divided into only one IRS, the optimal deployment location is relatively close to the APs for both the active and passive IRS. this is because it allows all the UEs to benefit from the IRS beamforming as much as possible, and maximizes the average ergodic throughput of all UEs. 
When divided into multiple IRSs, such as 4 IRSs, the optimal deployment locations for active and passive are different. For active IRSs, the IRSs can be placed farther away from the APs because the active IRSs have power amplification. However, for the passive IRSs, the optimal deployment location of the IRSs will not be very far away from the APs so as to increase the power reaching the IRSs.

\subsection{Network Coverage Ratio with Increasing Number of IRSs in a Wider Area}
In contrast to the preceding two subsections, the purpose of optimization here is to maximize the coverage ratio of the UEs. 
We substitute the objective function with the sum of coverage indicators in optimization problem (P1) and solve it using Gurobi. 
The main parameter settings are the same as the previous ones, while a larger area of 1080 m$\times$1600 m is considered. 
Moreover, the number of UEs is also expanded to 200, the length and width of each grid are 20 m and 7 m, and the number of selected candidate grid points is 176.
Note that at longer distances, the coverage capability of the AP is weakened, and it is more meaningful to introduce the IRS for coverage enhancement. 

\begin{figure}
    \centering
        \includegraphics[width=1\linewidth, trim=50 0 75 42,clip]{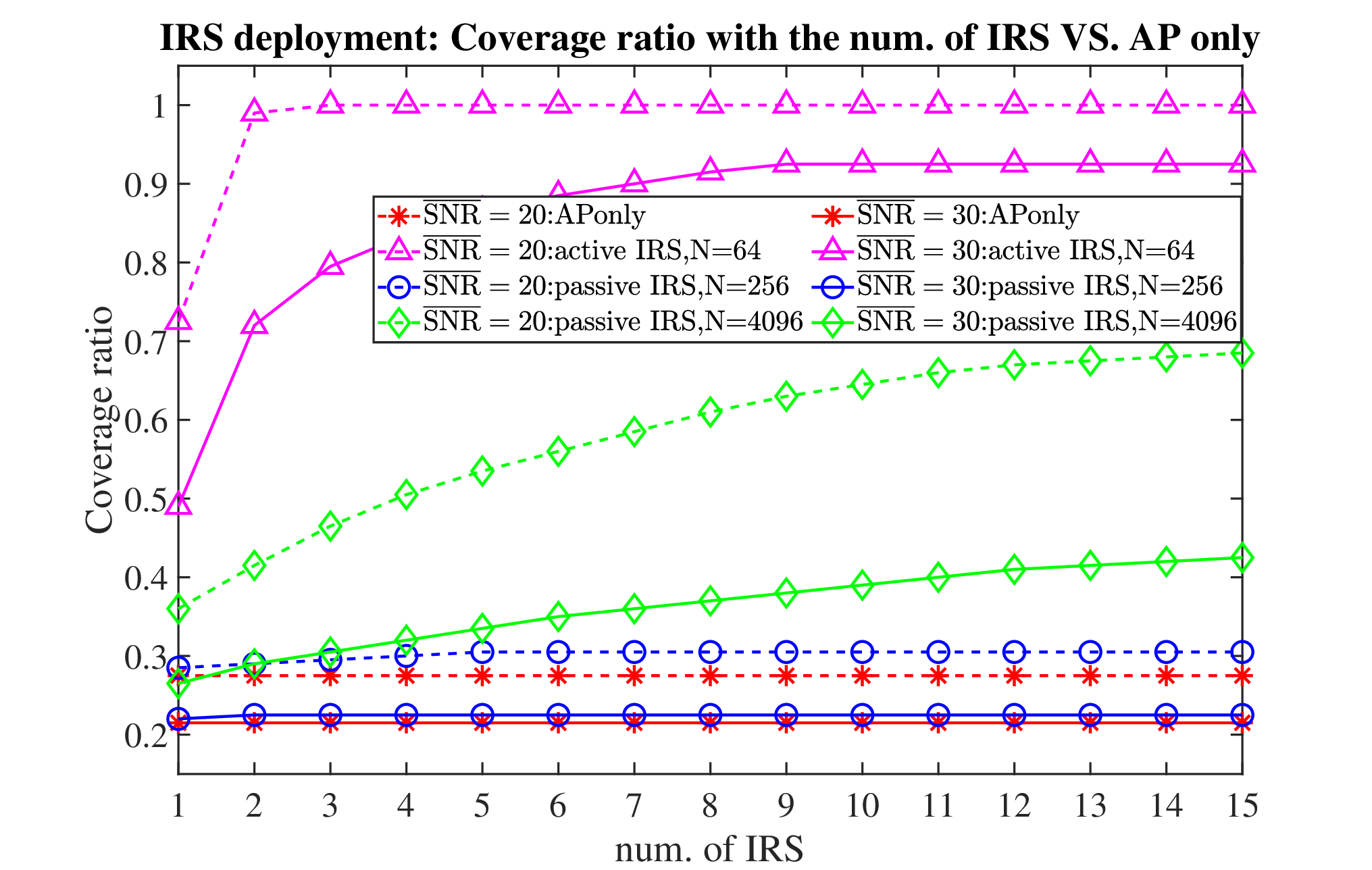}
        % \caption{Comparison experiment of active IRS, passive IRS, and no IRS deployment.\vspace{-2ex}}
        \caption{Comparison of network coverage ratio by different schemes in a wide area.\vspace{-2ex}}
        \label{IRS_deployment-coverage_ratio}
\end{figure}

In this experiments, we set the SNR thresholds as $\overline{\rm{SNR}}=20\,\rm{ dB}, 30\,\rm{ dB}$, respectively. The experimental results are shown in Fig. \ref{IRS_deployment-coverage_ratio}.
It is evident that the deployment of IRSs can enhance network coverage, irrespective of whether thresholds are set at 20 dB or 30 dB. However, active IRSs exhibit superior efficacy, achieving substantial coverage improvements with a smaller number of IRSs. In contrast, passive IRSs require a greater number and larger size to achieve comparable coverage enhancements.
Especially at high thresholds (i.e., $\overline{\rm{SNR}}=30\,\rm{dB}$), the coverage performance improvement due to passive IRSs is more constrained. This limitation arises because passive IRSs have an inherently weaker channel enhancement capability, which is insufficient to meet the demand of a higher threshold.

\section{Conclusion}
   In this paper, we first investigate the link-level performance of active or passive IRS taking into account the IRS ERP and AP antenna radiation pattern. Then the network-level coverage performance is evaluated/optimized in site-specific multi-building scenarios, by properly deploying multiple IRSs on candidate building facets to serve a given set of users or PoIs. The problem is reduced to an integer linear programming (ILP) based on given link-level metrics, which is then solved efficiently under moderate network sizes. 
    Numerical results confirm the impact of AP antenna/IRS element pattern on the link-level performance. In addition, it is found that active IRSs, though associated with higher hardware complexity and cost, significantly improve the site-specific network coverage performance in terms of average ergodic rate and fairness among the PoIs as well as the range of serving area, compared with passive IRSs that have a much larger number of elements.
In future work, we will investigate multi-AP configuration, inter-IRS reflection, interference, and resource scheduling.

\bibliography{IEEEabrv,bibliography}

%附录部分

% \begin{appendix}
\begin{appendices}

\end{appendices}

\end{document}